\documentclass[lengthcheck,twocolumn,showpacs,superscriptaddress]{revtex4-1} 

\usepackage{amssymb}
\usepackage{subfigure}
\usepackage{amsmath}

\usepackage{graphicx}
\usepackage{epstopdf}
\usepackage{color}
\usepackage{ulem}

\newcommand{\scl}{0.20} 

\begin{document}

\title{Radically Tunable Ultrafast Photonic Oscillators via Differential Pumping}

\date{\today}

\author{Yannis Kominis}
\affiliation{School of Applied Mathematical and Physical Science, National Technical University of Athens, Athens, Greece	}

\author{Anastassios Bountis}
\affiliation{Department of Mathematics, Nazarbayev University, Astana, Republic of Kazakhstan}

\author{Vassilios Kovanis}
\affiliation{Bradley Department of Electrical and Computer Engineering, Virginia Tech, Arlington, USA}.
\affiliation{School of Optics and Photonics - CREOL, The University of Central Florida, Orlando, USA}

\begin{abstract}
We present the controllability capabilities for the limit cycles of an extremely tunable photonic oscillator, consisting of two coupled semiconductor lasers. We show that this system supports stable limit cycles with frequencies ranging from a few to more than a hundred GHz that are characterized by a widely varying degree of asymmetry between the oscillations of the two lasers. These dyamical features are directly controllable via differential pumping as well as optical frequency detuning of the two lasers, suggesting a multi-functional oscillator for chip-scale radio-frequency photonics applications. 
\end{abstract}

\maketitle

Limit cycles are the fundamental ingredients of a wide variety of physical as well as man-made systems exhibiting characteristic self-sustained oscillations. Their existence is directly related to the interplay of two characteristic features that can be met in almost all realistic models, namely nonlinearity and dissipation. In contrast to oscillations of conservative nonlinear systems whose frequency is determined by the initial energy of the system, limit cycles have frequencies that are determined solely by the parameters of the system and often constitute global attractors to which the system evolves for any initial condition \cite{Strogatz_Book}. The existence of such robust  limit cycles renders them crucial for the life itself, since they often occur in chemical and biological rhymes such as circadian oscillations \cite{Pavlidis_Book,  Winfree_Book,  Kuramoto_Book, Strogatz_Sync}, whose frequencies are determined by enviromental physical parameters. On the other hand, the existence of such limit cycles in man-made systems is crucially important for key technological applications such as time or frequency references \cite{Udem_02, Hollberg_02} with their range of frequencies being controllable by the system parameters with significantly greater freedom in comparison to biological oscillators.

Currently, there is intense interest in various implementations of ultrafast reconfigurable oscillators in systems and functional devices for next generation Photonic Integrated Circuits (PIC) \cite{Bowers_18, Coldren_Book} and RF photonics applications \cite{Kovanis_14}. Due to the fact that single semiconductor lasers are not capable of supporting limit cycles, configurations based on optically coupled lasers have been proposed and intensively studied for more than four decades \cite{Winful_Book}.  
In this context, Optically Injected Lasers (OIL) corresponding to a one-way coupling in a master-slave configuration [Fig. 1(a)] have been shown to support relatively tunable limit cycles \cite{OIL_1, OIL_2, OIL_3, OIL_4}; however, the need for a bulky optical isolator prevents their on-chip integration \cite{Coldren_Book} and significantly restricts their applications. The utilization of strong mutual coupling, corresponding to complicated configurations where a single electric field mode is amplified by two gain blocks, has been shown to result to a gain-lever mechanism [Fig. 1(b)] allowing for significant bandwidth-enhancing \cite{Gain-Lever_1, Gain-Lever_2}. The case of evanecently coupled diode lasers is shown to be the most promising for photonic integration \cite{Coldren_Book} and also capable of supporting stable limit cycles \cite{Winful_88, Winful&Wang_88, Winful_92}. However, for more than three decades the tunability capabilities of such structures have been limited due to the consideration of arrays of identical coupled lasers [Fig. 1(c)]. It is worth mentioning that for laser separation distances larger than those typically occurring in PICs or in cases where delays are intentionally introduced, such systems are described by delay-differential equations having a rich set of dynamical features \cite{LangKobayashi_80, Fischer_02, Mandel_03, Lenstra_05}. 

The introduction of topological characteristics in coupled lasers in terms of differential pumping and frequency detuning between the lasers [Fig. 1(d)] enables the on-chip implementation of a large set of key functionalities such as reconfigurable beam forming and steering \cite{Choquette_13}, coherence tuning and enhanced phase-locking \cite{Choquette_15, Kominis_17b, Adams_17}, localized syncrhonization \cite{Kovanis_97, Kominis_17b}, enhanced bandwidth and tailored modulation response \cite{Choquette_17mod, Kominis_19} as well as existence of exceptional points allowing for ultra-sensitivity \cite{Choquette_17, Choquette_18, Kominis_17a, Kominis_18}. 

In this work we consider the fundamental non-Hermitian optical meta-molecule consisting of two mutually coupled and differentially pumped semiconductor lasers as the basic reconfigurable oscillator element of a photonic integrated circuit exhibiting an extreme frequency tunability spanning over 100 GHz and controlled by minute changes of the electrically injected differential pumping. The latter is shown to control not only the frequency but also the shape of the underlying limit cycle providing a remarkable flexibility for the RF properties of the emited light beam.     

The time evolution of the electric fields and the number densities of two evanescently coupled diode lasers is governed by the following coupled single-mode rate equations for the amplitude of their normalized electric fields $E_1, E_2$, their phase difference $\theta$ and the normalized excess carrier densities $N_1, N_2$:
\begin{eqnarray}
 \dot{E_1}&=&E_1N_1-\Lambda E_2\sin\theta \nonumber \\
 \dot{E_2}&=&E_2N_2+\Lambda E_1\sin\theta \nonumber \\
 \dot{\theta}&=&\Delta -\alpha(N_2-N_1)+\Lambda\left(E_1/E_2-E_2/E_1\right)\cos\theta   \label{pair} \\
T\dot{N_1}&=&P_1- N_1-(1+2 N_1)E_1^2 \nonumber \\
T\dot{N_2}&=&P_2- N_2-(1+2 N_2)E_2^2 \nonumber
\end{eqnarray}
Here $\alpha$ is the linewidth enhancement factor, $\Lambda$ is the normalized coupling constant, $P_i$ is the normalized excess pumping rate, $\Delta=\omega_2-\omega_1$ is the cavity optical detuning, $T$ is the ratio of carrier to photon lifetimes, and $t$ is the time normalized to the photon lifetime $\tau_p$ \cite{Winful&Wang_88, Choquette_13}. \

The phase-locked states of the system (\ref{pair}), are obtained by setting the time derivatives of the system equal to zero and their stability is determined by the eigenvalues of the Jacobian of the linearized system. For the case of zero detuning ($\Delta =0$) and symmetric pumping ($P_1=P_2=P_0$), two phase-locked states are known analytically: $E_1=E_2=\sqrt{P_0}$, $N_1=N_2=0$ and $\theta=0,\pi$. The in-phase state ($\theta_s=0$) is stable for $\Lambda>\alpha P_0 /(1+2P_0)$ whereas the out-of-phase state ($\theta_s=\pi$) is stable for $\Lambda<(1+2P_0)/2\alpha T$ \cite{Winful&Wang_88}. Under general conditions of nonzero detuning and differential pumping the phase-locked states can be asymmetric and are characterized by three parameters $(E_0, \rho, \theta_s)$, namely the value of the electric field amplitude of the first laser $E_0 \equiv E_1$, the field amplitude ratio $\rho \equiv E_2/E_1$, and the phase difference $(\theta_s)$  \cite{Kominis_17a}. The appropriate values of the system parameter set consisting of the pumping rates $P_i$ and detuning $\Delta$ can be determined as functions of the solution characteristics $(E_0, \rho, \theta_s)$ as follows:
\begin{eqnarray}
 \Delta&=&-\alpha \Lambda\sin\theta_s\left(\rho^{-1}+\rho\right)-\Lambda\cos\theta_s\left(\rho^{-1}-\rho\right)  \label{D_eq}\\
 P_1&=&E_0^2+(1+2E_0^2)\Lambda\rho\sin\theta_s  \nonumber\\
 P_2&=&\rho^2E_0^2-(1+2\rho^2 E_0^2) \Lambda\rho^{-1}\sin\theta_s  \label{P_eq}
\end{eqnarray}
and the corresponding steady-state values of $N_i$ are:
\begin{eqnarray}
N_1&=&\Lambda \rho \sin\theta_s \nonumber \\
N_2&=&-\Lambda\rho^{-1}\sin\theta_s \label{Z_eq}
\end{eqnarray} 
The above equations clearly show that $P_i$ and $\Delta$ can be chosen to yield a phase-locked state of arbitrary amplitude asymmetry $\rho$, phase difference $\theta_s$ and reference amplitude $E_0$. In the case of zero detuning ($\Delta=0$) between the coupled lasers, the phase difference is restricted as 
\begin{equation}
\theta_s=s\pi + \tan^{-1}\left[\frac{1}{\alpha} \frac{\rho^2-1}{\rho^2+1}\right], \hspace{2em} s=0,1 \label{theta_rho} \\ 
\end{equation}
whereas in the case of symmetrically pumped lasers ($P_1=P_2=P_0$) the reference amplitude $E_0$ is fixed as
\begin{equation}
E_0^2 = \frac{ \Lambda \sin\theta_s (\rho^2+1)}{\rho\left[(\rho^2-1)-4 \Lambda \rho \sin\theta_s\right]} \label{X0_rho}
\end{equation}
and the common pumping rate is $P_0=E_0^2+(1+2E_0^2) \Lambda \rho \sin\theta_s$. It is worth mentioning that experimental constraints may restrict the simultaneous control of the two pumping rates and the frequency detuning. These are related to the fact that the magnitudes of the injection currents not only change the pump parameters but also vary the cavity resonance frequency through ohmic heating and refractive index temperature dependency. However, as we show in the following, for the general case of unequal pumping and nonzero detuning, there is always enough flexibility in the parameter space of the system for the appropriate selection of $P_1, P_2$ and $\Delta$ in order to support a limit cycle with desired frequency, even when taking into account such experimental constraints.

%
The stability of the phase-locked states is determined by the eigenvalues of the Jacobian of the dynamical system (\ref{pair}). When a pair of complex conjugate eigenvalues crosses the imaginary axis, the system undergoes Hopf bifurcations that give rise to stable limit cycles corresponding to undamped relaxation oscillations characterized by asymmetric synchronized oscillations of the electric fields, that can have different mean values and amplitudes \cite{Kovanis_97, Kominis_17b}. In addition to the parameters of the corresponding phase-locked states from which they have emerged, the limit cycles are characterized by their frequencies $f_H$ and  ratios $R$ of the peak-to-peak electric field oscillation amplitudes that can be determined as the imaginary part of the eigenvalues with zero real part at the Hopf bifurcation point and the ratio of the first two components of the corresponding eigenvectors, respectively.
In what follows we focus on a system with parameters values corresponding to recent experiments on coherently coupled phased photonic crystal vertical cavity lasers \cite{Choquette_17mod} as shown in Table I.

\begin{table}
\centering
 \begin{tabular}{|c c c |} 
 \hline
Symbol & Parameter & Value  \\ 
 \hline \hline
 $\alpha$ & Linewidth enhancement factor & $4$  \\ 
 \hline
 $\tau_c$ & Carrier lifetime & $2$ns \\
\hline
 $\tau_p$ & Photon lifetime & $2 \times 10^{-3}$ns \\
 \hline
 $T$ & Ratio of carrier to photon lifetime & $1000$  \\ 
 \hline
 $\Lambda$ & Normalized coupling coefficient & $10^{-4} - 10^0$ \\ 
 \hline
\end{tabular}
\caption{Realistic parameter values for coherently coupled phased photonic crystal vertical cavity lasers \cite{Choquette_17mod}.}
\end{table}

The case of zero optical detuning $\Delta=0$ and equal pumping $P_1=P_2=P_0$ is depicted in Fig. 2, where the Hopf frequency $(f_H)$ and the ratio of amplitude oscillations $(R)$ are shown as functions of the corresponding phase-locked state asymmetry $\rho$ and the coupling parameter $\Lambda$. For this case the phase difference $(\theta_s)$ and electric field amplitude $E_0$ are restricted by Eqs. (\ref{theta_rho}) and (\ref{X0_rho}) respectively. The frequencies of the emerging limit cycles can vary from sub-GHz to more than a hundred GHz as shown in Fig. 2(a). Limit cycles with high-frequencies exist for strong coupling and emerge from phase-locked states with $\rho$ close to unity with similar oscillation amplitudes for the two lasers $R\simeq 1$ as shown in Fig. 2(b). Lower coupling values correspond to highly asymmetric limit cycles with much lower (even sub-GHz) frequencies. The Hopf frequencies can be  significantly higher (two orders of magnitude) than the free-running relaxation frequencies of an isolated quantum well laser with the same pumping rate given as $f_{rel}=(1/2\pi)\sqrt{2 P_0/T}$. Characteristic cases of stable limit cycles are shown in Fig. 3.\

For the case of zero detuning $(\Delta=0)$ and unequal pumping $P_1 \neq P_2$ the electric field amplitude can be arbitrary and the Hopf frequency as a function of $\rho$ and $\Lambda$ as well as $P_1-P_2$ and $\Lambda$ is depicted in Figs. 4(a) and 4(b). Limit cycles with high frequencies emerge again from quite symmetric phase-locked states $(\rho \simeq 1)$ for strong coupling. In contrast to the previous case of equal pumping, for smaller values of the coupling constant for which we have lower frequencies, not only the corresponding phase locked state can be very asymmetric, but also the oscillation amplitudes can be quite different $R >> 1$ as shown in Fig. 4(c). In such cases, the laser with the higher electric field mean value undergoes oscillations with much smaller amplitude as shown in Fig. 5.\

 For the most general case of non-zero detuning $\Delta \neq 0$, the phase-locked states from which the stable limit cycles emmerge can have arbitrary phases $(\theta_s)$ and electric field amplitude $(E_0)$. As shown in Fig. 6, in such case, extremely high values of Hopf frequencies (hundreds of GHz) can be obtained even for weaker coupling (such as $\log \Lambda = -2.2$) for phase differences $\theta_s \lesssim \pi$. The ratio of amplitude oscillations $R$ varies drastically as $\theta_s$ varies in the interval $(\pi/2, \pi)$ whereas for $\theta_s<\pi/2$ the phase-locked states undergo saddle-node bifurcations and do not give rise to limit cycles. As shown in Fig. 7, the electric field amplitude oscillations of the two lasers can have very different characteristics with either the one with the lower or the one with the higher mean value having the larger oscillation amplitude. In the limiting case of very high oscillation frequencies, the electric field of the first laser appears almost constant at a high value whereas the second one oscillates with a significant amplitude around a much lower mean value corresponding to $\rho << 1$ and $R >> 1$. \
 
 It is worth noting that this limiting case resembles the case of an injection locked laser, although in our system there is no need for an optical isolator since the differential pumping allows for the support of strongly asymmetric phase-locked states of the system of two mutually coupled lasers. In fact, by setting $Y=E_2, Z=N_2, \theta_s=\pi/2-\psi, \Lambda E_0=\eta$ and $\Delta = \Omega$, in the limit of $\rho \rightarrow 0$, the system of Eq. (\ref{pair}) is decoupled as 
\begin{eqnarray}
 \dot{E_1}&=&E_1N_1 \nonumber \\
 T\dot{N_1}&=&P_1- N_1-(1+2 N_1)E_1^2 \nonumber 
\end{eqnarray}
with steady state solution $N_1=0$ and $E_1=\sqrt{P_1}$ and

\begin{eqnarray}
 \dot{Y}&=&YZ+\eta\sin\psi \nonumber \\
 \dot{\psi}&=&-\Omega + \alpha Z-\frac{\eta}{Y} \cos\psi   \\
T\dot{Z}&=&P_2- Z-(1+2 Z)Y^2 \nonumber
\end{eqnarray}
which is the well-known injection model \cite{Erneux_book}. In such case, the Hopf frequency is given by
\begin{equation}
 f_H=(1/2\pi) \sqrt{\omega_{rel}^2+\eta^2}
\end{equation}
where $\omega_{rel}$ is the free running relaxation frequency of the slave laser and it scales linearly with the injection rate $\eta$ for large injection rates \cite{OIL_2}. 
 
In our case of differentially pumped mutually coupled lasers, a simple asymptotic analysis in terms of the small parameter $\epsilon=1/T$ shows that the eigenvalues of the Jacobian, for strongly asymmetric $(\rho << 1)$ phase-locked states with phase difference $\theta_s \rightarrow \pi$, can be approximated as
\begin{equation}
 \lambda=\frac{\Lambda\rho}{2}\pm i\frac{\Lambda}{\rho}.
\end{equation}
Therefore, the frequencies of the limit cycles emerging through Hopf bifurcations scale as 
\begin{equation}
 f_H=\frac{1}{2\pi} \frac{\Lambda}{\rho}
\end{equation}
which clearly reveals the role of the diffential pumping and the resulting asymmetry as an enabler for ultra-high frequency tuning even for moderate coupling strength, in accordance to Fig. 6(a).

In conclusion, we have demonstrated the radical tunability of a photonic oscillator consisting of two mutually coupled semiconductor lasers. The extreme frequency tuning spans over 3 orders of magnitude going beyond 100 GHz and along with the distribution of the electric field amplitude, they can be controlled by minute changes of the electrically injected differential pumping. This extreme tunability is enabled by the asymmetric topology of the system and is remarkable for any physical or man-made system supporting self-sustained oscillations. These properties suggest this photonic meta-molecule as a fundamental reconfigurable ultrafast oscillator for photonic integrated circuits for radio frequency photonics applications, optical communications as well as chip-scale frequency synthesis and sensing.

The authors acknowledge stimulating discussions on laser dynamics with Thomas Erneux, Athanasios Gavrielides and Luke Lester. This research is partly supported by two ORAU grants entitled "Taming Chimeras to Achieve the Superradiant Emitter" and ''Dissecting the Collective Dynamics of Arrays of Superconducting Circuits and Quantum Metamaterials'', funded by Nazarbayev University. Funding from MES RK state-targeted program BR05236454 is also acknowledged.

\begin{figure*}[pt]
  \begin{center}
  \subfigure[]{\scalebox{\scl}{\includegraphics{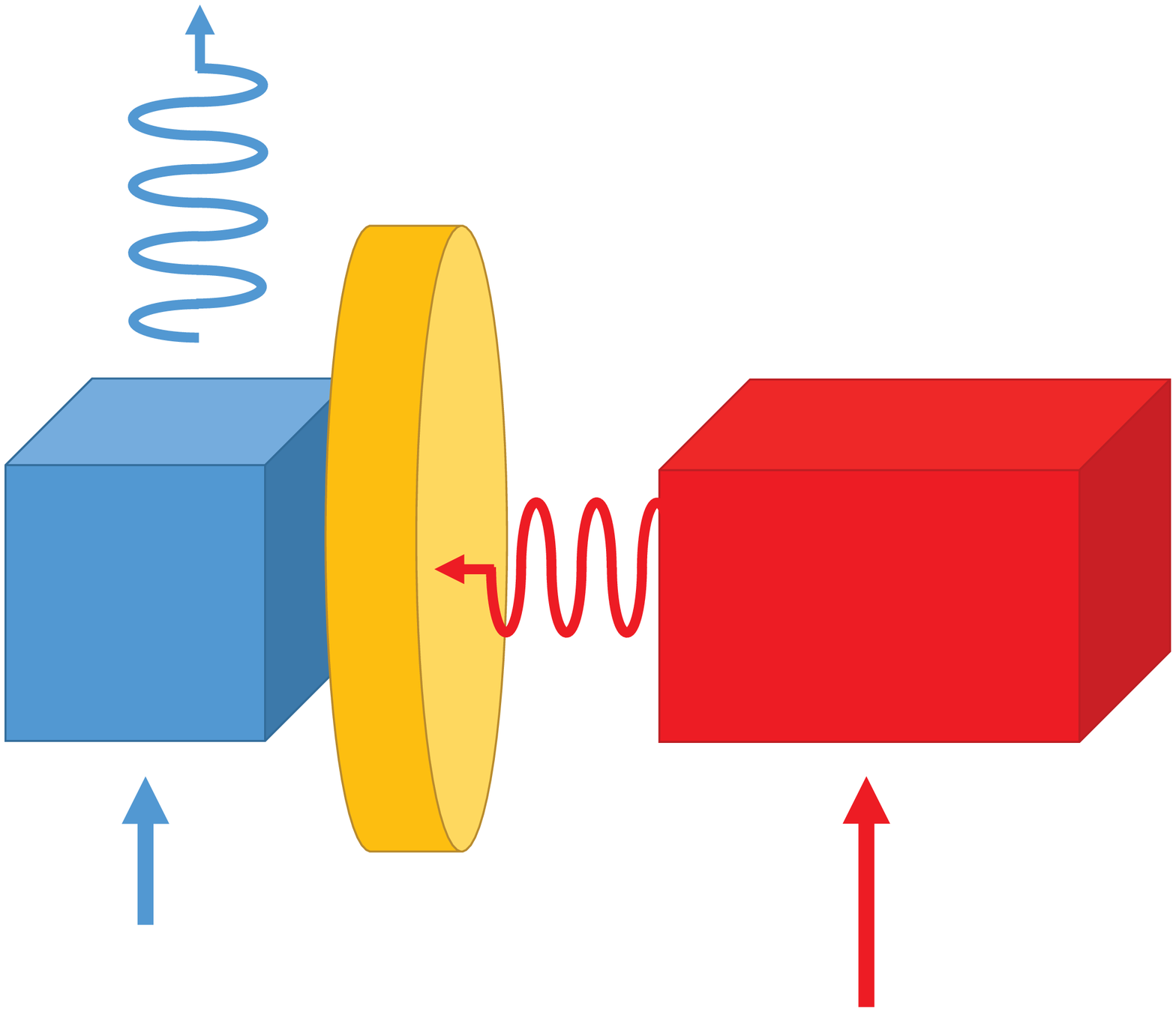}}}
  \subfigure[]{\scalebox{\scl}{\includegraphics{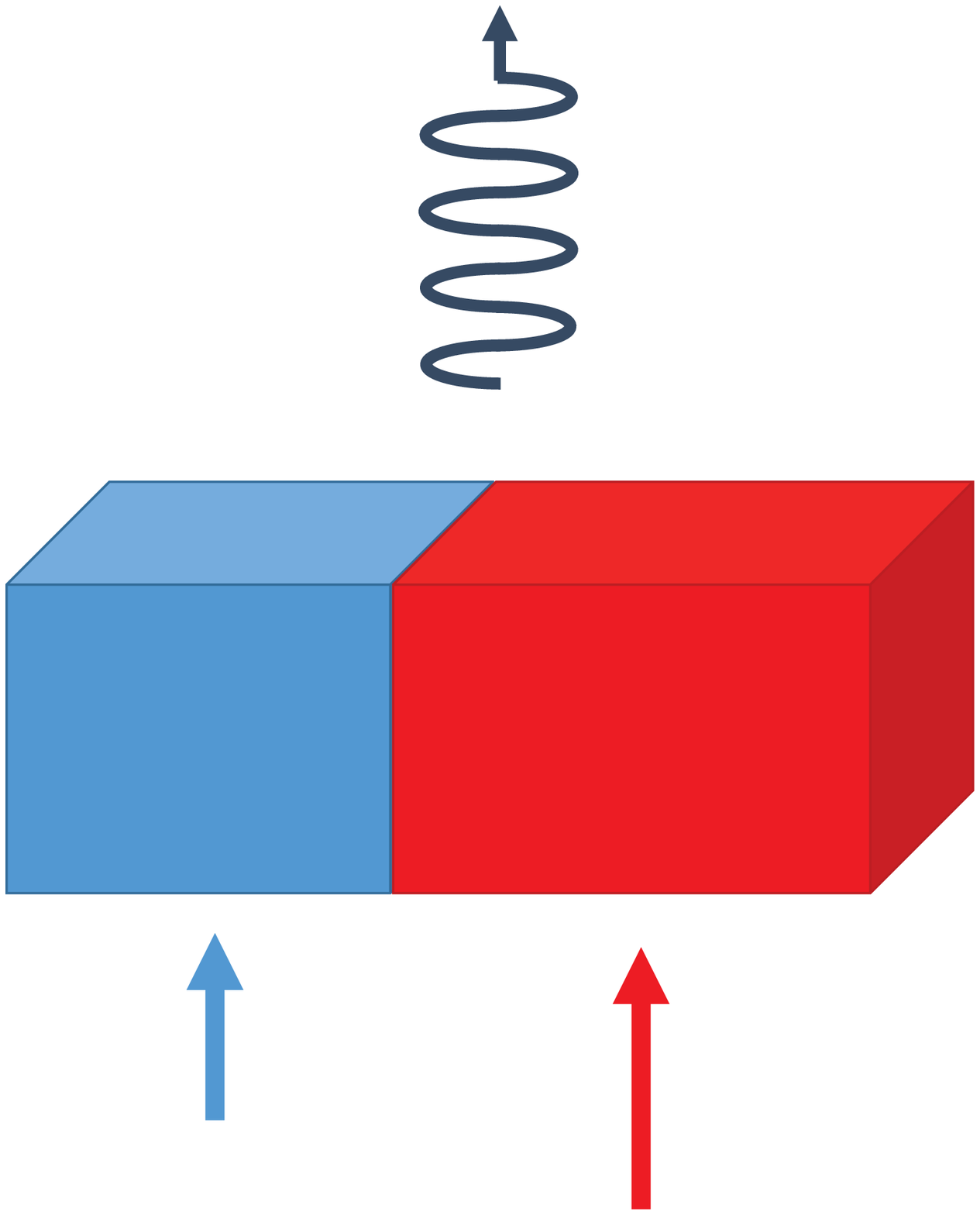}}}\\
  \subfigure[]{\scalebox{\scl}{\includegraphics{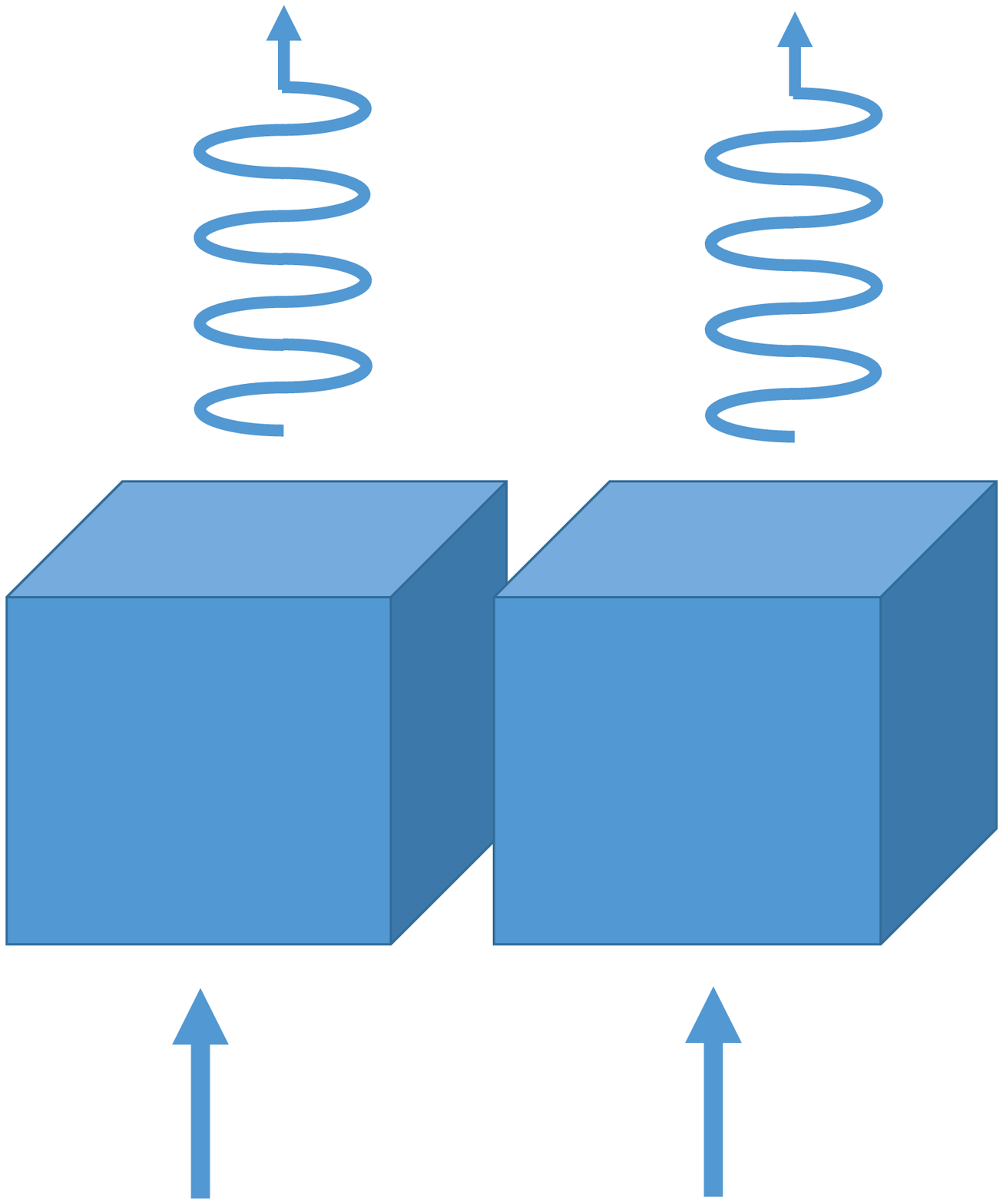}}}
  \subfigure[]{\scalebox{\scl}{\includegraphics{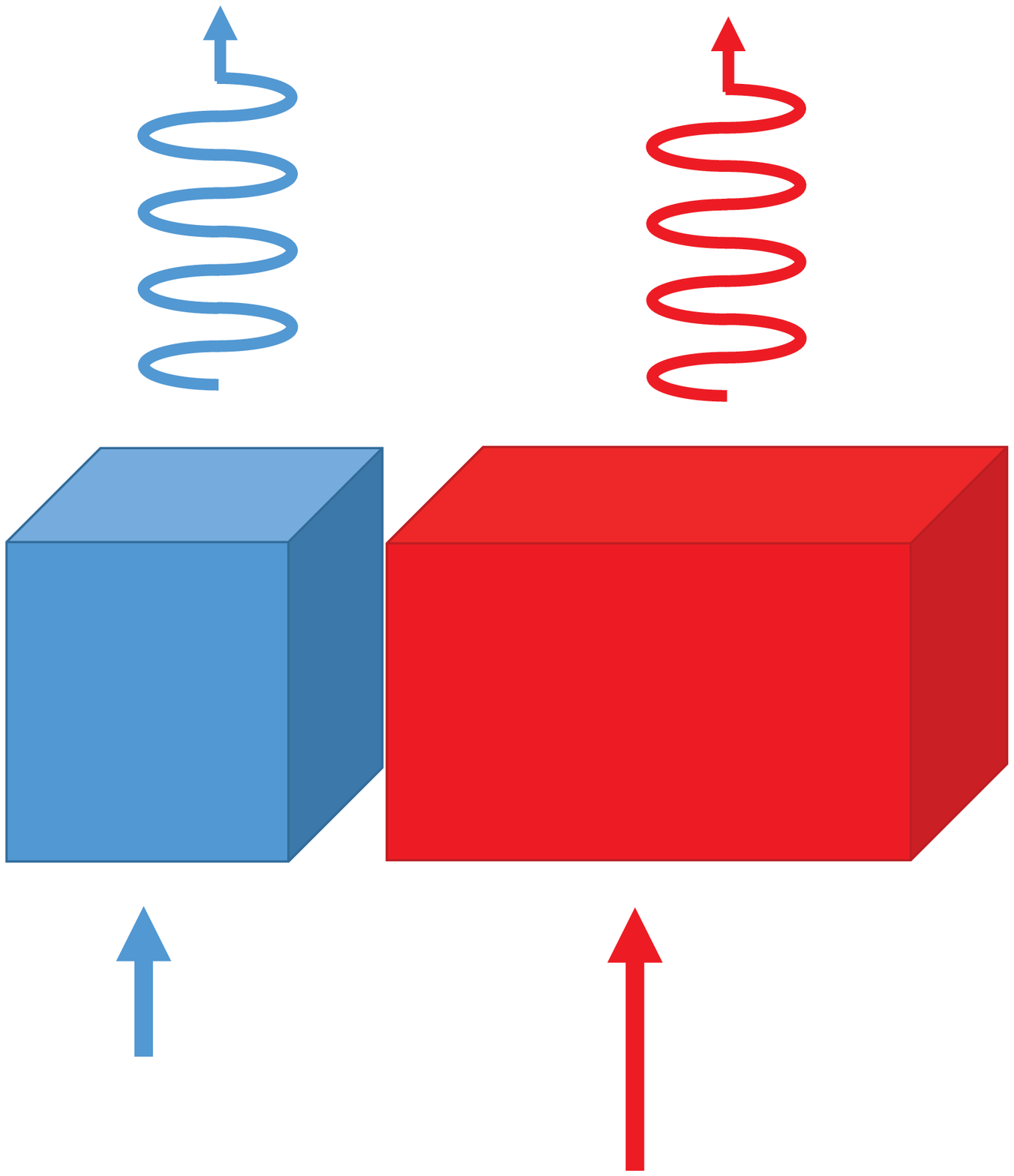}}}
  \caption{Configurations of two Coupled Lasers varying in terms of: symmetry of the two gain blocks, electrically injected pumping and output electric field(s): (a) Optically Injected Laser, (b) Gain-Lever Configuration, (c) Evanecently Coupled Identical Lasers, (d) Evenecently Coupled Differentially Pumped Dissimilar Lasers. Different block and arrow sizes/colors denote different gain blocks and pumping rates, respectively.}  
  \end{center}
\end{figure*}

\begin{figure*}[pt]
  \begin{center}
  \subfigure[]{\scalebox{\scl}{\includegraphics{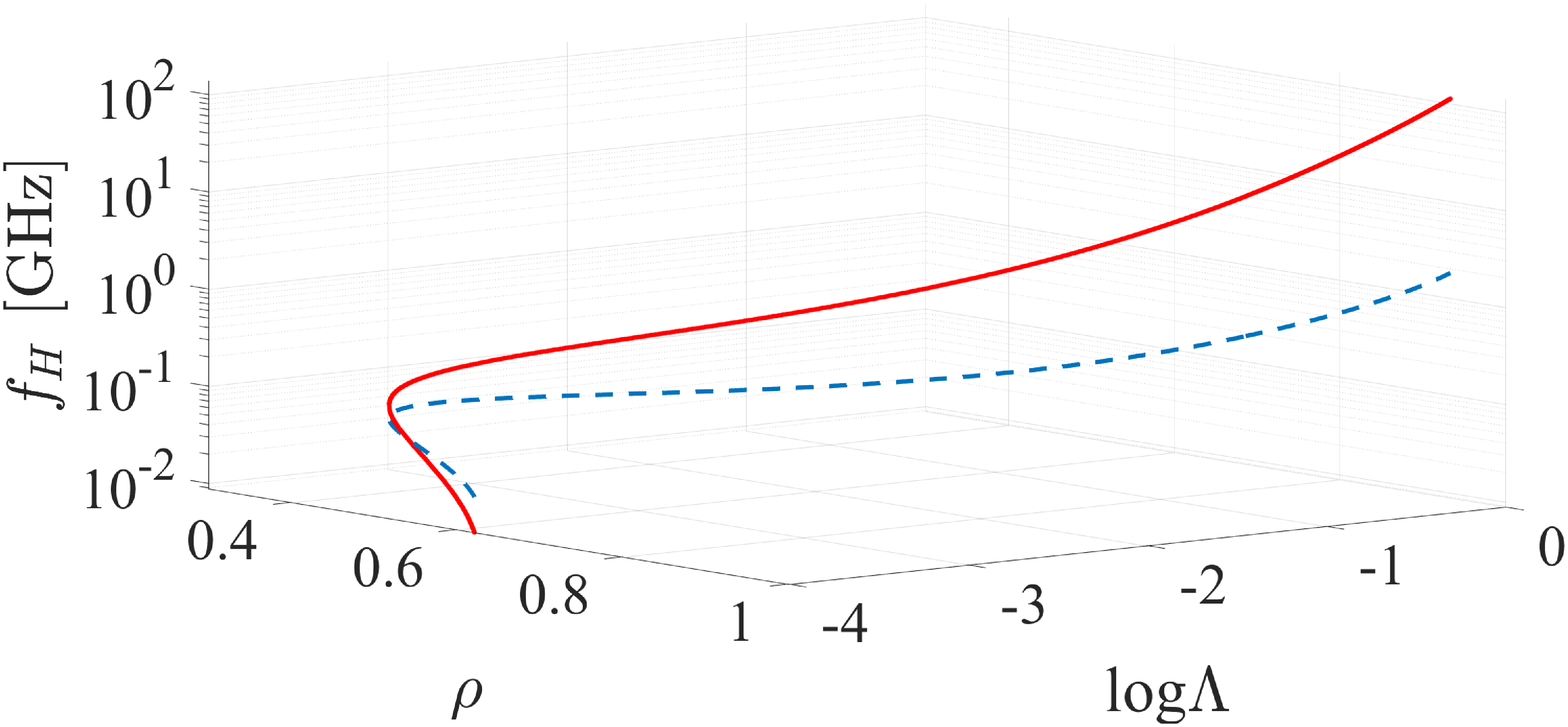}}}
  \subfigure[]{\scalebox{\scl}{\includegraphics{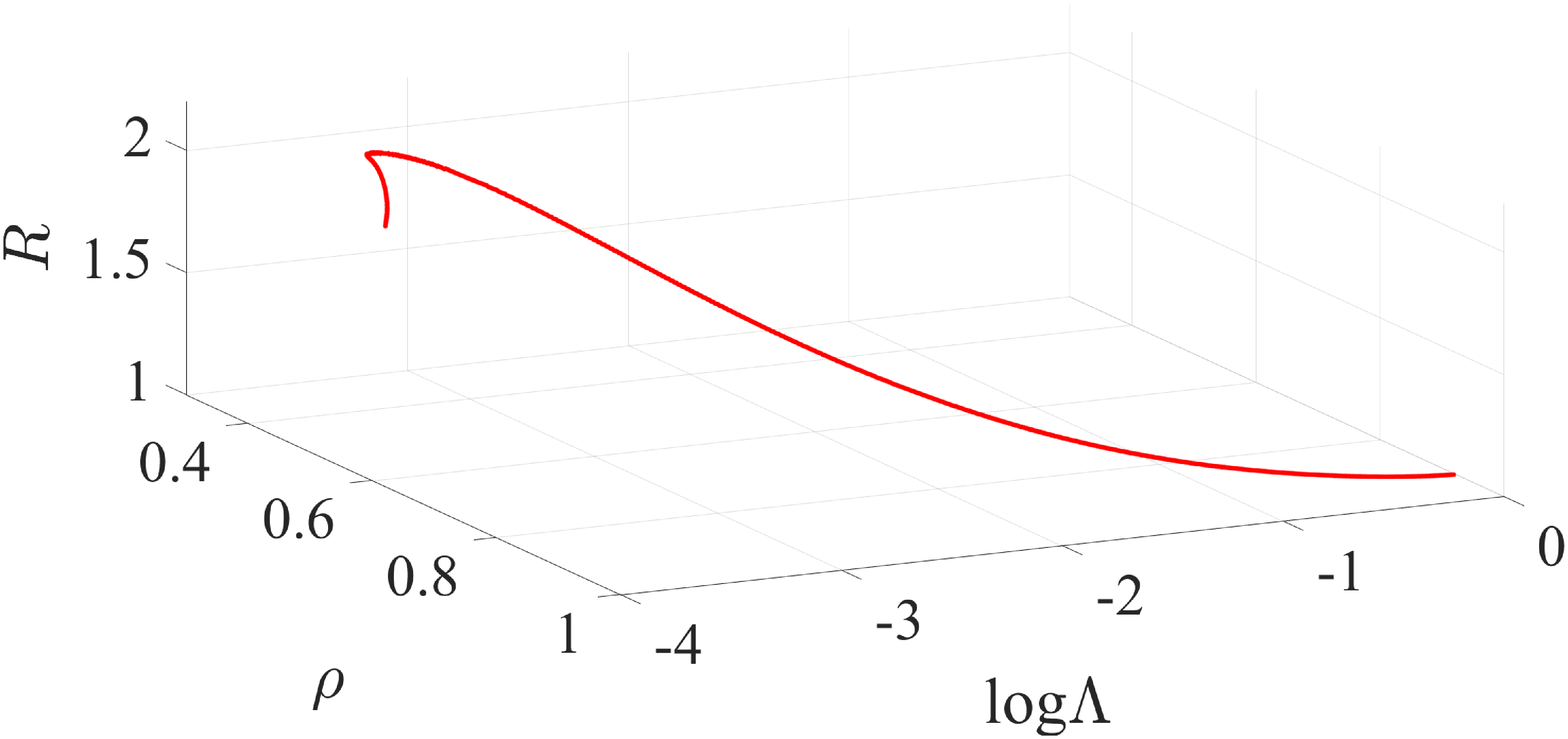}}}
  \caption{Case of zero detuning ($\Delta=0$) and equal pumping ($P_1=P_2=P_0$): Hopf frequencies $f_H$ (a) and oscillation amplitude ratios $R$ (b) of stable limit cycles that are emanated from the asymmetric phase locked states as functions of the asymmetry parameter $(\rho)$ corresponding to the electric fields amplitude ratio and the coupling strength $\Lambda$. The dashed line in (a) depicts the free running relaxation frequency $f_{rel}=(1/2\pi)\sqrt{2 P_0/T}$. The Hopf frequency remains close to $f_{rel}$ for weak coupling and radically increases at more than two orders of magnitude higher values for stronger coupling for which the corresponding phase-locked state is more symmetric $(\rho\simeq 1)$ (for example: $f_H=100$ GHz and $f_{rel}=1.71$ GHz at $\log\Lambda=-0.2$). This family of limit cycles is drastically different from previously well known limit cycles emanating from symmetric states \cite{Winful&Wang_88}.}  
  \end{center}
\end{figure*}

\begin{figure*}[pt]
 \begin{center}
  \subfigure[]{\scalebox{\scl}{\includegraphics{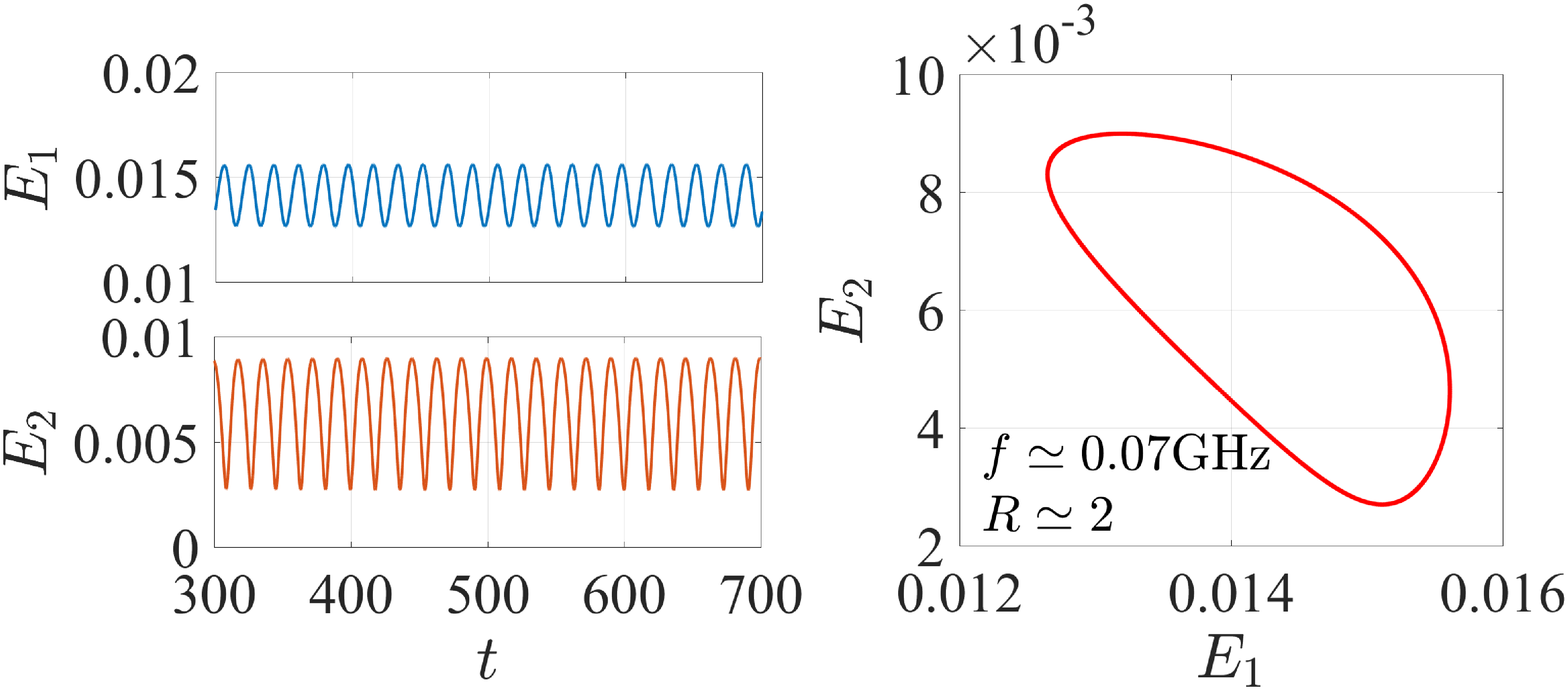}}}
  \subfigure[]{\scalebox{\scl}{\includegraphics{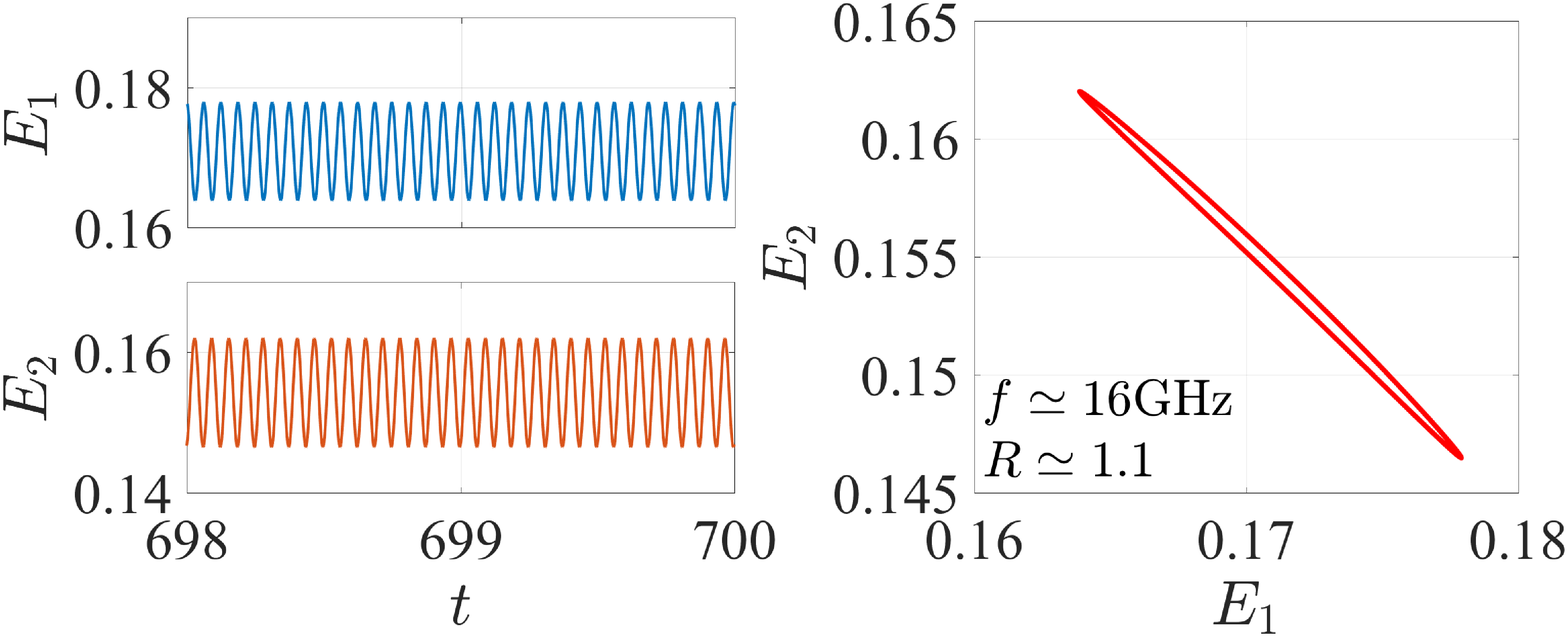}}}
  \caption{Characteristic stable limit cycles for weak (a) and strong (b) coupling for the case of zero detuning ($\Delta=0$) and equal pumping ($P_1=P_2=P_0$). Coupling strength and asymmetry parameters: (a) $(\log\Lambda,\rho)=(-3.5,0.39)$, (b) $(\log\Lambda,\rho)=(-1,0.89)$.}
  \end{center}
\end{figure*}

\begin{figure*}[pt]
  \begin{center}
  \subfigure[]{\scalebox{\scl}{\includegraphics{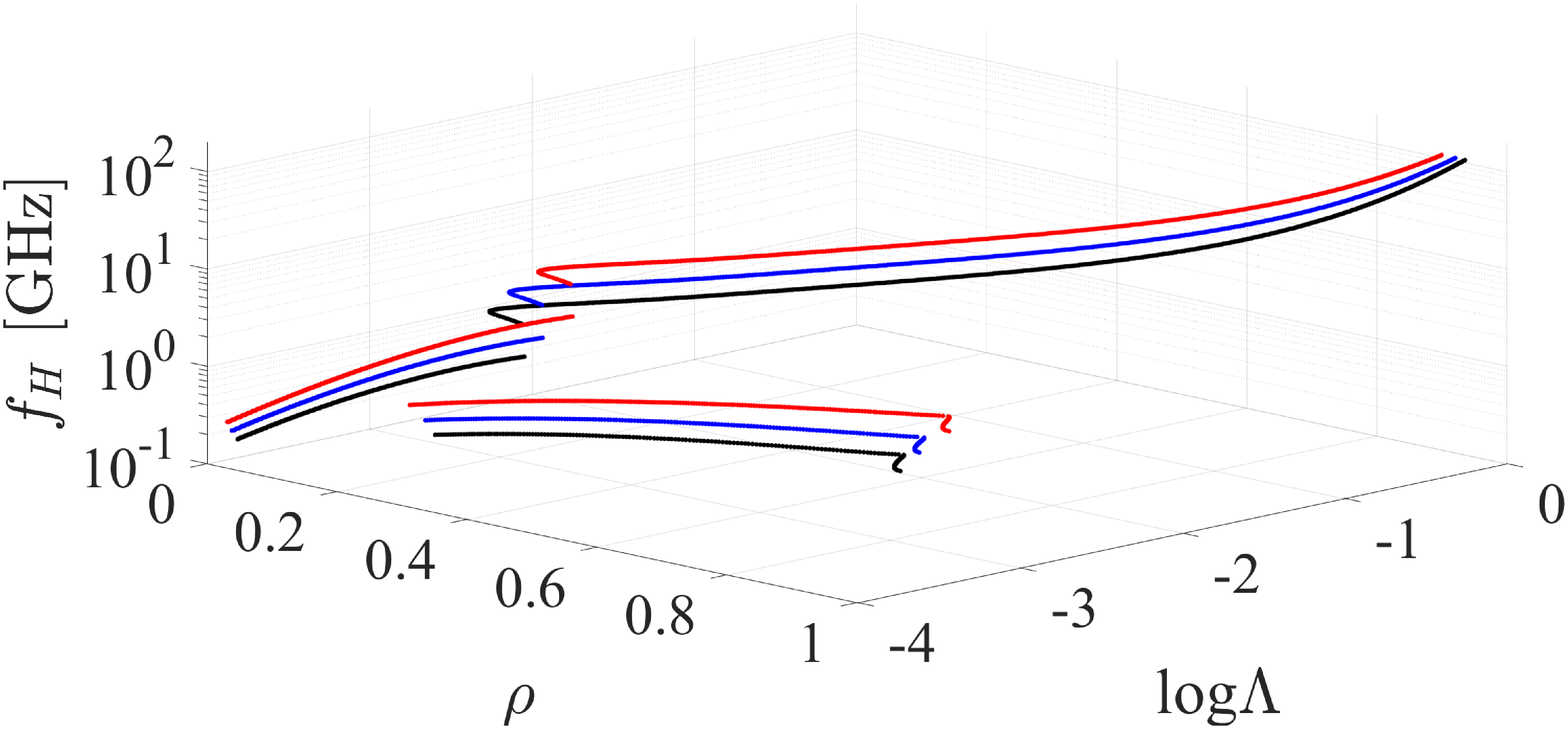}}}
  \subfigure[]{\scalebox{\scl}{\includegraphics{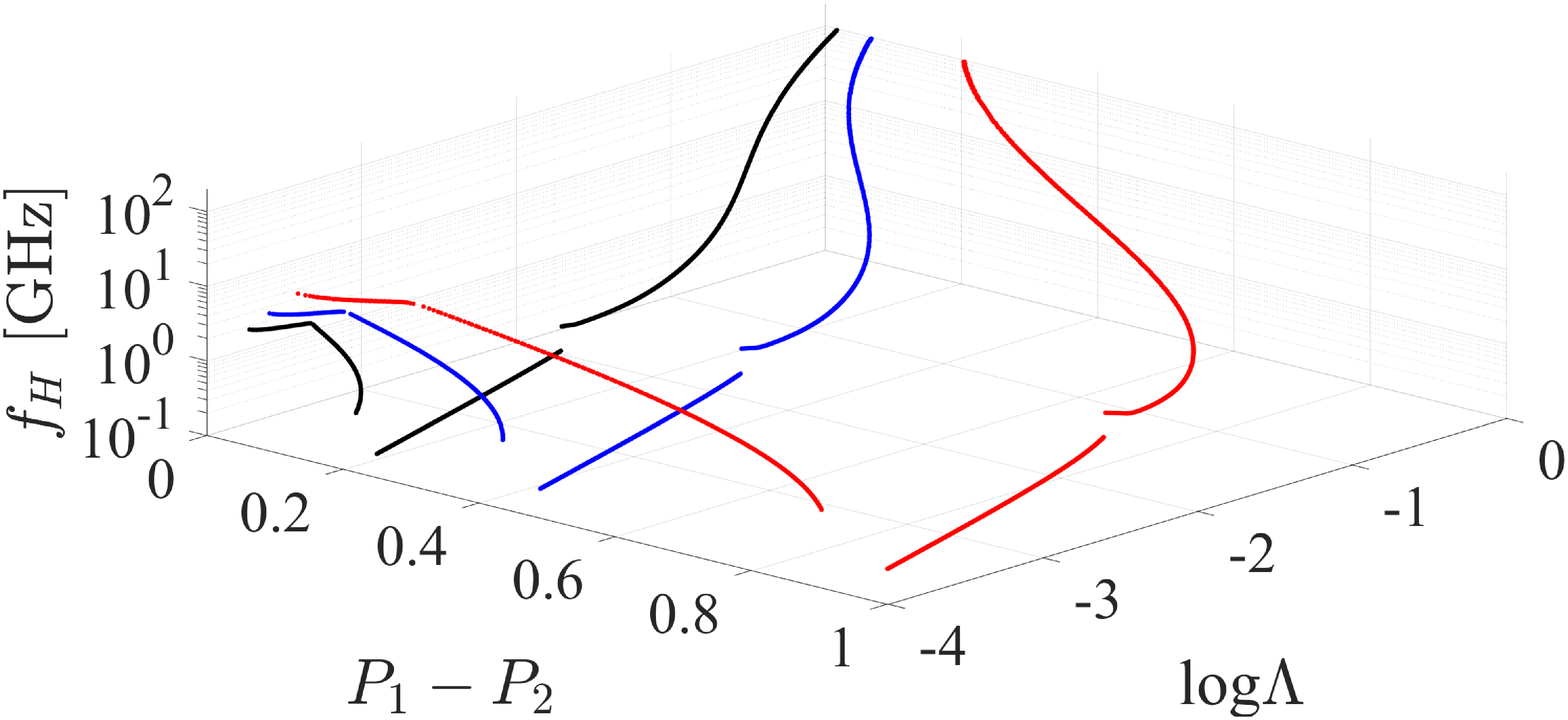}}}\\
  \subfigure[]{\scalebox{\scl}{\includegraphics{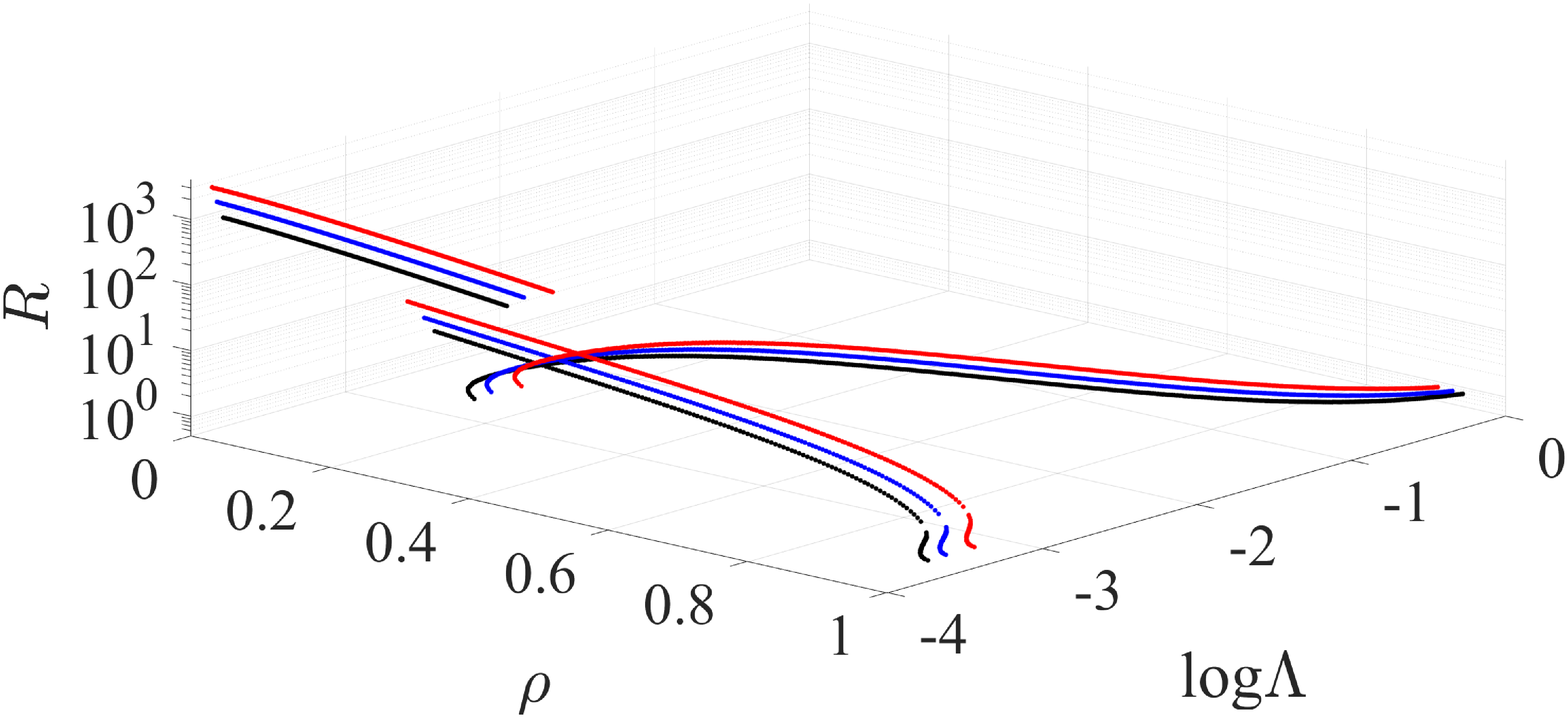}}}
  \caption{Case of zero detuning ($\Delta=0$) and unequal pumping ($P_1 \neq P_2=P_0$): Hopf frequncies $f_H$ of stable limit cycles as functions the asymmetry of the corresponding phase locked state $(\rho)$ and the coupling strength $(\Lambda)$ (a) as well as of $\rho$ and $P_1-P_2$ (b). Oscillation amplitude ratios $(R)$ as function of $\rho$ and $\Lambda$ (c). The corresponding  phase-locked states are characterized by $s=1$ and $E_0=0.5,0.7,1$ (black, blue,red). The differential pumping allows for arbitrary electric field amplitudes as given by $E_0$. Although there is no significant difference in the dependence $f_H$ and $R$ on the asymmetry parameter $\rho$ and the coupling strength $\Lambda$, different values for $E_0$ allow for smaller or larger pumping differences with similar effects with respect to $f_H$ and $R$.} 
  \end{center}
\end{figure*}

\begin{figure*}[pt]
  \begin{center}
  \subfigure[]{\scalebox{\scl}{\includegraphics{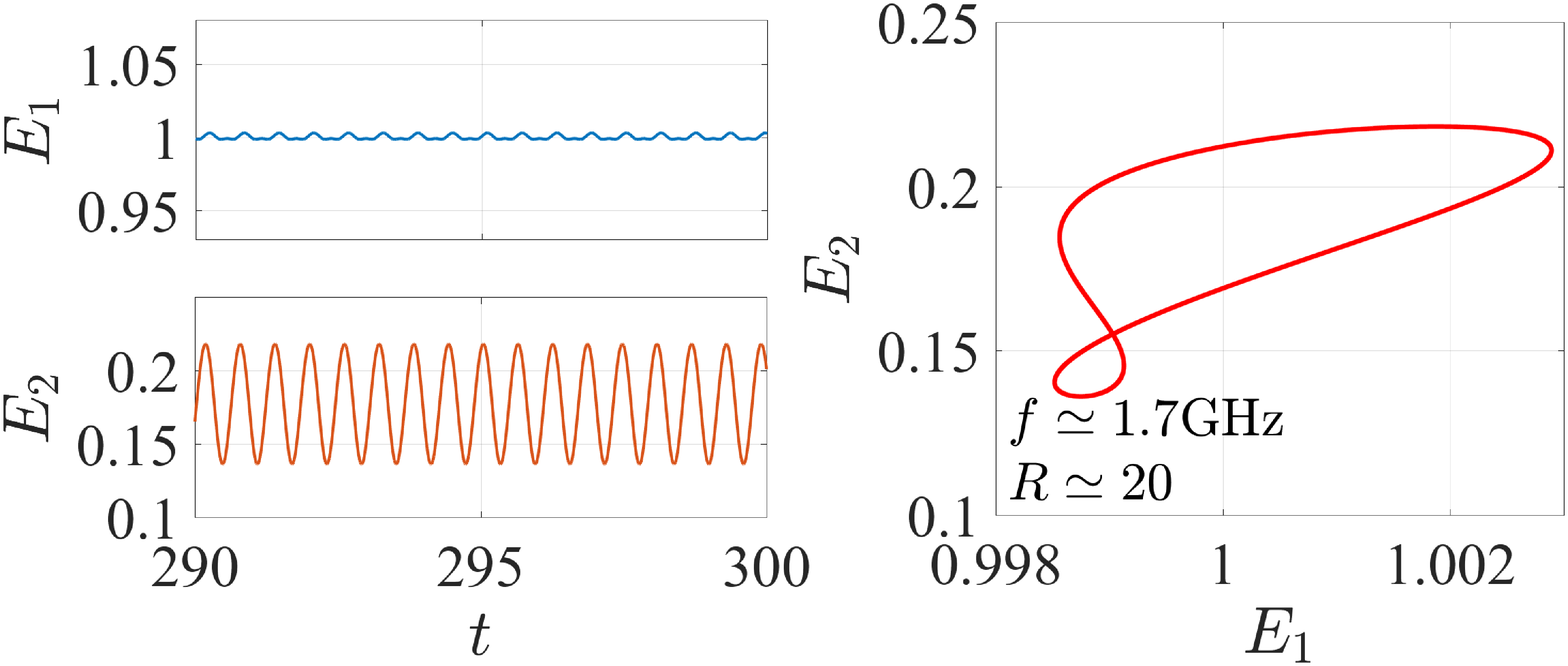}}}
  \subfigure[]{\scalebox{\scl}{\includegraphics{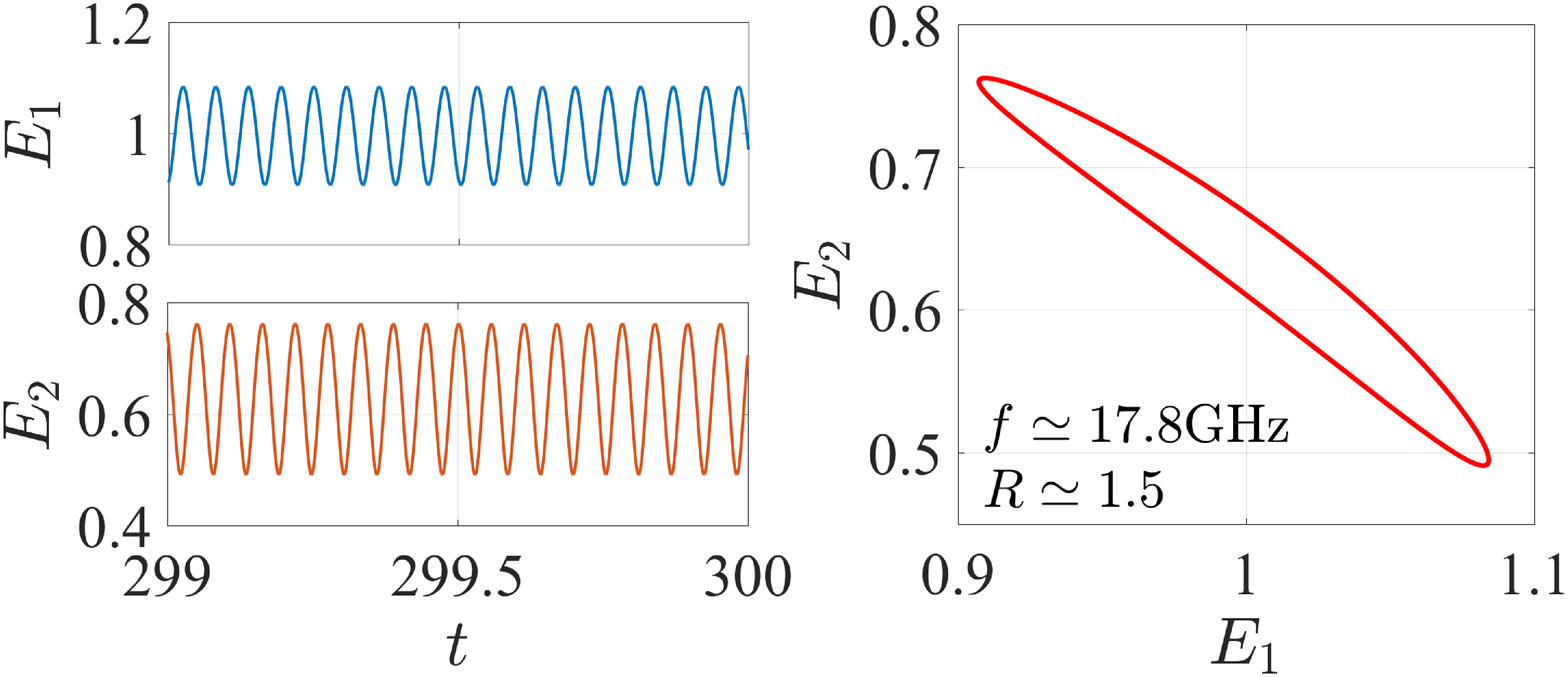}}}
  \caption{Characteristic stable limit cycles for weak (a) and strong (b) coupling for the case of zero detuning ($\Delta=0$) and unequal pumping ($P_1\neq P_2=P_0$). The corresponding  phase-locked state is characterized by $s=1$ and $E_0=1$. Coupling strength and asymmetry parameters: (a) $(\log\Lambda,\rho)=(-2.5,0.18)$, (b) $(\log\Lambda,\rho)=(-1,0.64)$.}
  \end{center}
\end{figure*}

\begin{figure*}[pt]
  \begin{center}
  \subfigure[]{\scalebox{\scl}{\includegraphics{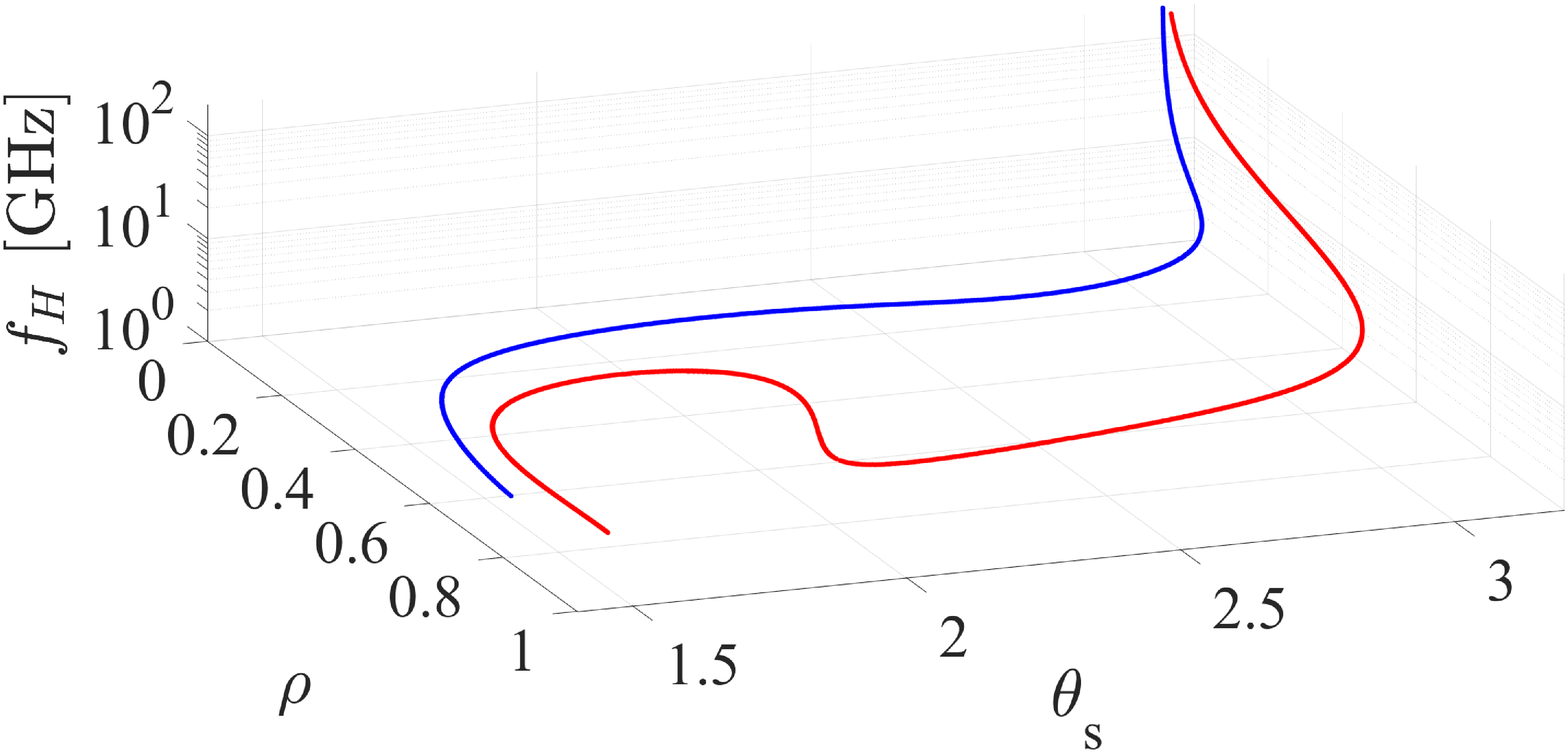}}}
  \subfigure[]{\scalebox{\scl}{\includegraphics{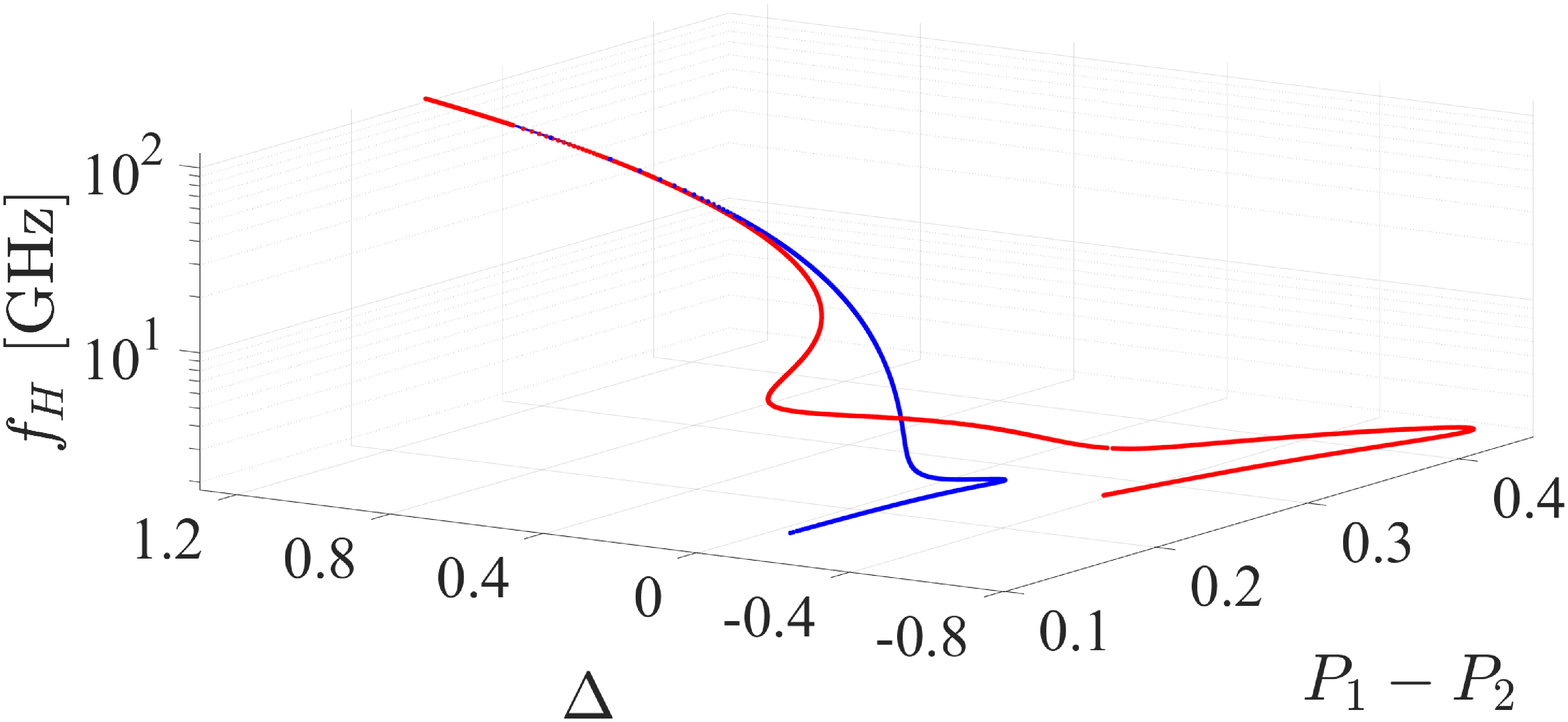}}}\\
  \subfigure[]{\scalebox{\scl}{\includegraphics{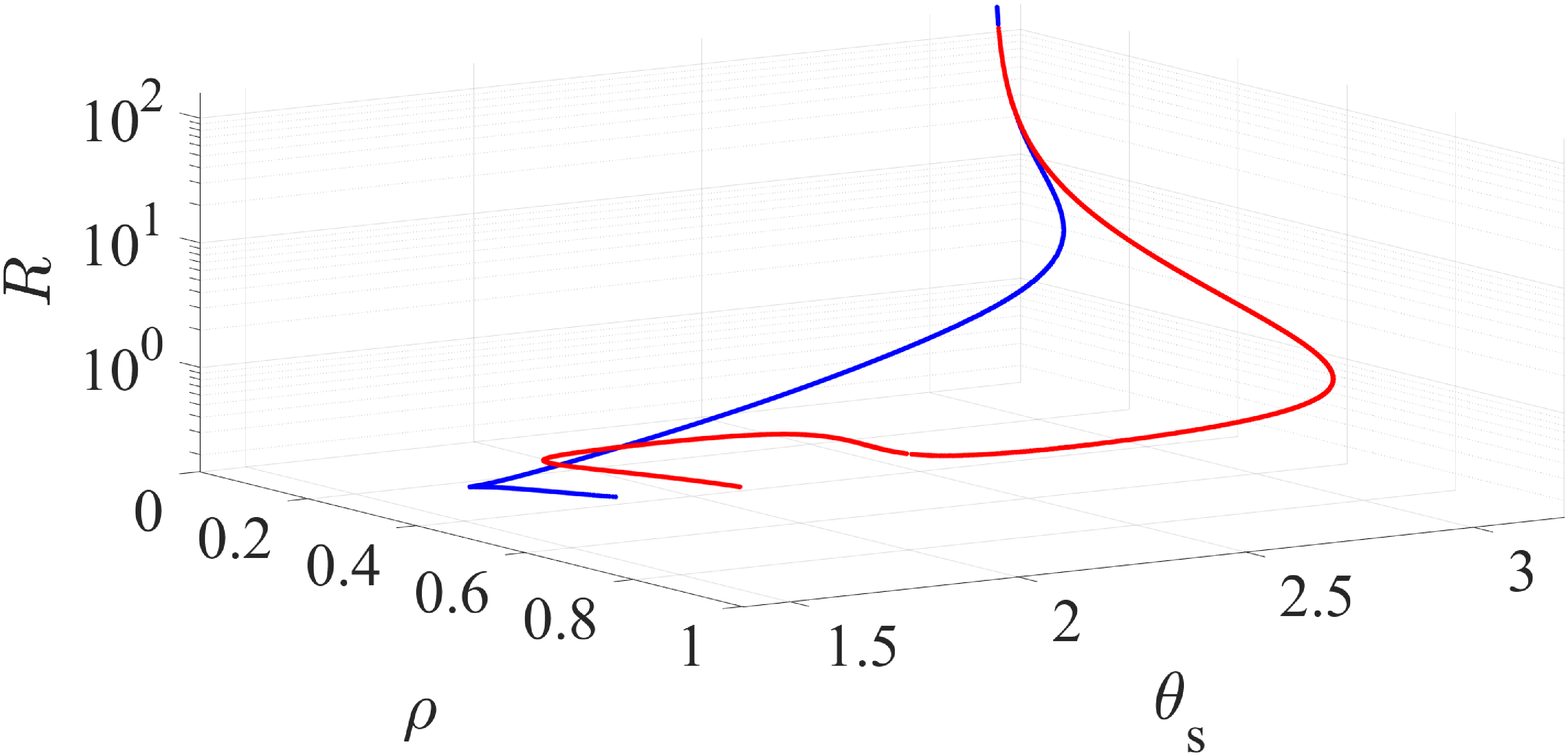}}}
  \caption{Case of nonzero detuning ($\Delta \neq 0$) and unequal pumping ($P_1 \neq P_2$): Hopf frequencies $f_H$ of stable limit cycles as functions of the asymmetry $(\rho)$ and the phase difference $(\theta_s)$ of the corresponding phase locked state (a) as well as of the pumping difference $P_1-P_2$ and detuning $\Delta$ (b). Oscillation amplitude ratios $(R)$ as function of $\rho$ and $\theta_s$ (c). The electric field amplitude of the first laser in the corresponding phase-locked state is $E_0=0.5$ and the coupling constant is $\log\Lambda=-2.2,-1.2$ (blue, red). In contrast to cases with zero detuning, the non-zero detuning allows for extremely high values of Hopf frequency at intermediate values of coupling strength as well as enhanced controllability of the ratio of the electric field oscillation amplitudes $R$ of the two lasers. }
  \end{center}
\end{figure*}

\begin{figure*}[pt]
  \begin{center}
  \subfigure[]{\scalebox{\scl}{\includegraphics{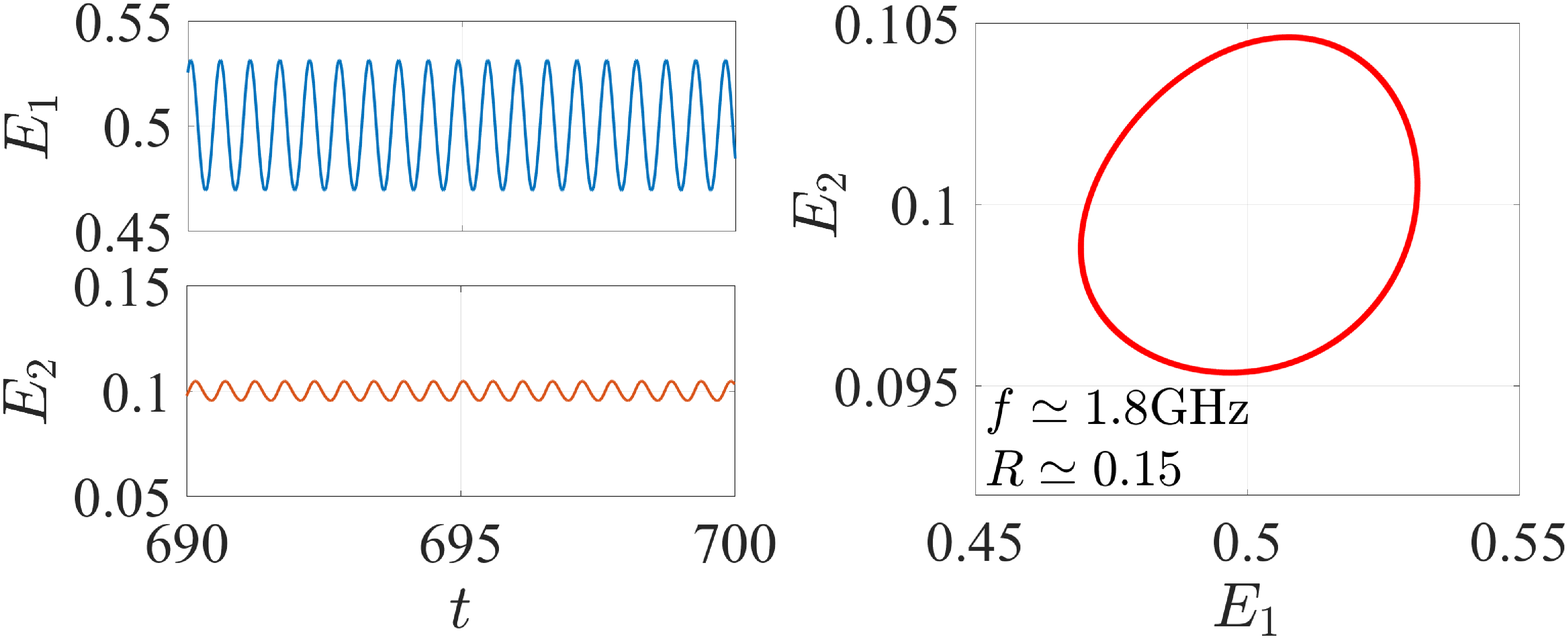}}}
  \subfigure[]{\scalebox{\scl}{\includegraphics{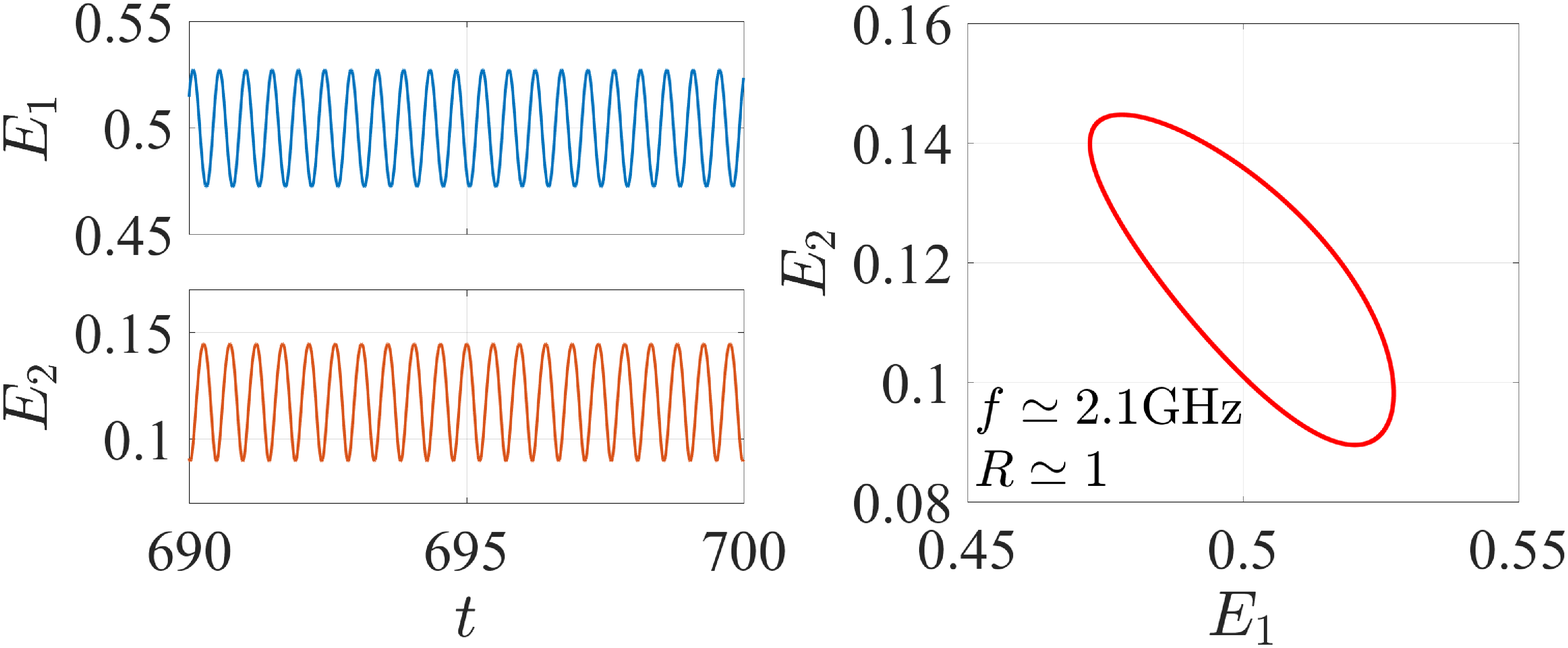}}}\\
  \subfigure[]{\scalebox{\scl}{\includegraphics{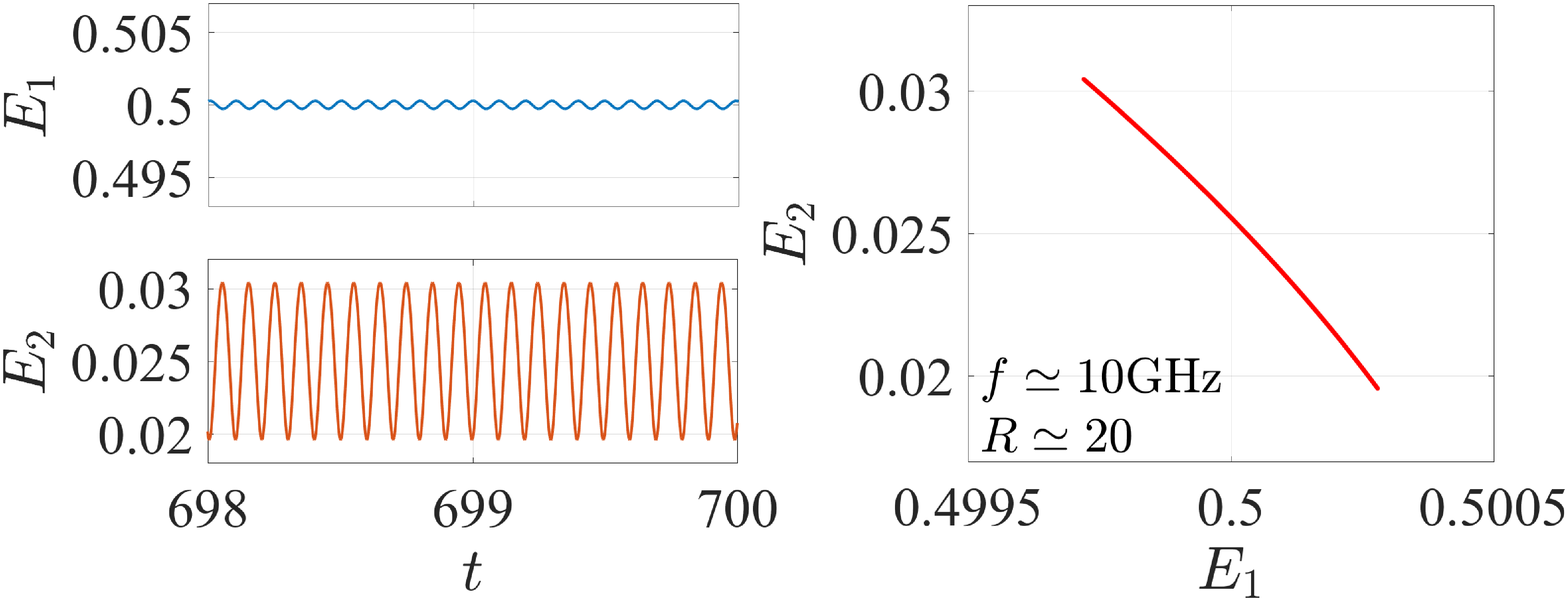}}}
  \subfigure[]{\scalebox{\scl}{\includegraphics{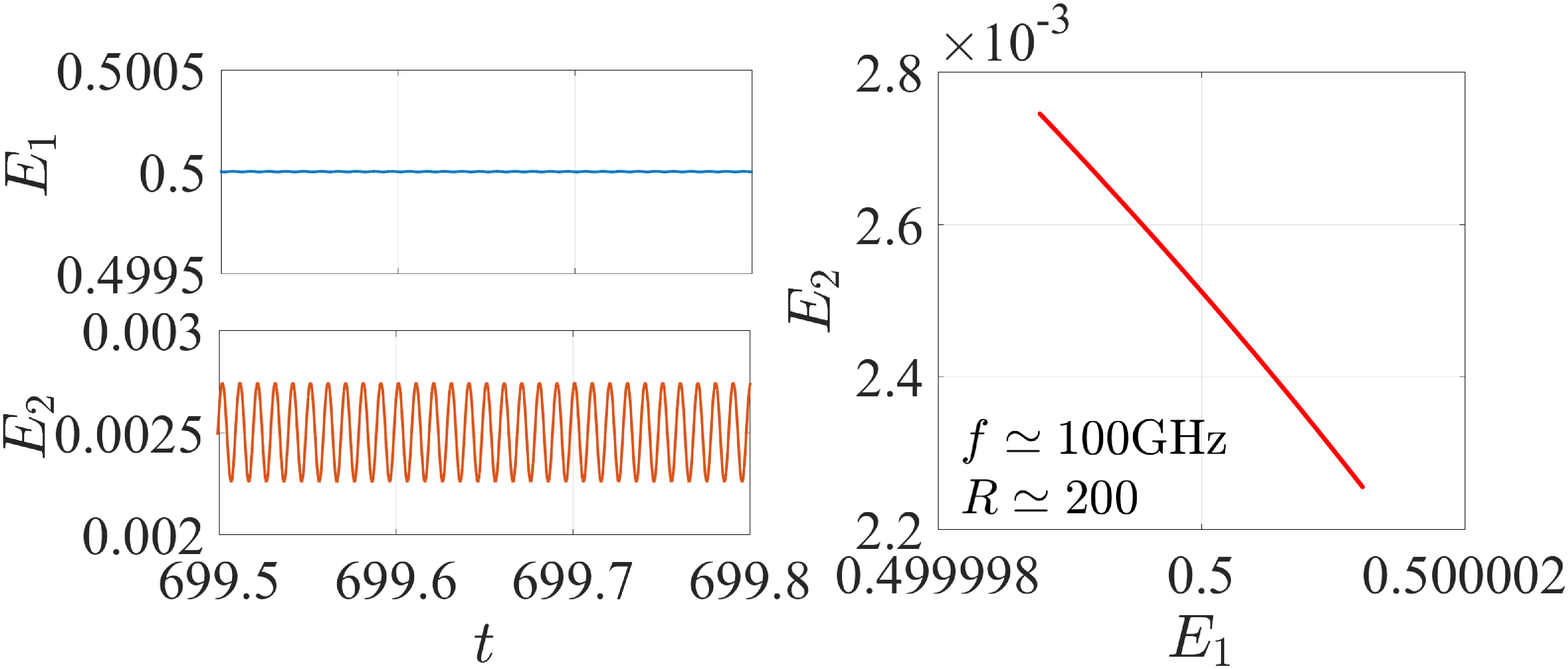}}}
  \caption{Characteristic stable limit cycles for the case of nonzero detuning ($\Delta \neq 0$) and unequal pumping ($P_1\neq P_2$). The corresponding  phase-locked state are characterized by $E_0=0.5$ and different phase differences $\theta_s$ and the coupling constant is $\log \Lambda=-2.2$. Phase difference and asymmetry parameters: (a) $(\rho, \theta_s)=(0.20,1.89)$, (b) $(\rho, \theta_s)=(0.24,2.75)$, (c) $(\rho, \theta_s)=(0.50,3.14)$, (d) $(\rho, \theta_s)=(0.005,3.14)$.} 
  \end{center}
\end{figure*}


\begin{thebibliography}{99}
\bibitem{Strogatz_Book} S. Strogatz, {\it Nonlinear Dynamics and Chaos}, CRC Press (2018).
\bibitem{Pavlidis_Book} T. Pavlidis, {\it Biological Oscillators: Their Mathematical Analysis}, Academic Press (1973).
\bibitem{Winfree_Book}  A.T. Winfree, {\it The Geometry of Biological Time}, Springer-Verlag (1980).
\bibitem{Kuramoto_Book} Y. Kuramoto, {\it Chemical Oscillations, Waves, and Turbulence}, Springer-Verlag (1984).
\bibitem{Strogatz_Sync} S. Strogatz, {\it Sync: How Order Emerges from Chaos in the Universe, Nature, and Daily Life}, Hyperion (2003).

\bibitem{Udem_02} T. Udem, R. Holzwarth, and T.W. Hanch, ``Optical frequency metrology,'' Nature \textbf{146}, 233-237 (2002).
\bibitem{Hollberg_02} J. Kitching,  S. Knappe, and L. Hollberg, ``Miniature vapor-cell atomic-frequency references,'' Appl. Phys. Lett., \textbf{81}, 553-555 (2002).

\bibitem{Bowers_18} T. Komljenovic, D. Huang, P. Pintus, M.A. Tran, M.L. Davenport, and J.E. Bowers, ``Photonic integrated circuits using heterogeneous integration on silicon,'' Proceedings of the IEEE,  \textbf{106}, 2246 - 2257 (2018).

\bibitem{Coldren_Book} L.A. Coldren, S.W. Corzine, and M.L. Masanovic, ``Diode Lasers and Photonic Integrated Circuits 2nd Ed.,'' Wiley (2012).

\bibitem{Kovanis_14} T.B. Simpson, J.M. Liu, M. AlMulla, N.G. Usechak, and V. Kovanis, ``Limit-Cycle Dynamics with Reduced Sensitivity to Perturbations,'' Phys. Rev. Lett. \textbf{112}, 023901 (2014).

\bibitem{Winful_Book} H.G. Winful and R.K. Defreez, ``Dynamics of coherent semiconductor laser arrays,'' (Diode Laser Arrays, Eds. D. Botez and D.R. Scifres), Cambridge University Press (1994).

\bibitem{OIL_1} R. Lang, ``Injection Locking Properties of a Semiconductor Laser,'' IEEE J. Quant. Electron. \textbf{18}, 976-983 (1982).

\bibitem{OIL_2} T.B. Simpson, J.M. Liu, A. Gavrielidies, V. Kovanis, and P.M. Alsing, ``Period-doubling cascades and chaos in a semiconductor laser with optical injection,'' Phys. Rev. A \textbf{51}, 4181-4185 (1995).

\bibitem{OIL_3} T.B. Simpson, J.M. Liu, and A. Gavrielides, ``Small-signal analysis of modulation characteristics in a semiconductor laser subject to strong optical injection,'' IEEE J. Quant. Electron. \textbf{32}, 1456-1468 (1996).

\bibitem{OIL_4} T.B. Simpson, J.-M. Liu, M. AlMulla, N.G. Usechak, and V. Kovanis, ``Tunable Oscillations in Optically Injected Semiconductor Lasers With Reduced Sensitivity to Perturbations,'' J. Light. Techn. \textbf{32}, 3749-3758 (2014).

\bibitem{Gain-Lever_1} K.J. Vahala, M.A. Newkirk, and T.R. Chen, ``The optical gain lever: A novel gain mechanism in the direct modulation of quantum well semiconductor-lasers,'' Appl. Phys. Lett. \textbf{54}, 2506-2508 (1989).

\bibitem{Gain-Lever_2} Y. Li, N.A. Naderi, V. Kovanis, and L.F. Lester, ``Enhancing the 3-dB bandwidth via the gain-lever effect in
quantum-dot lasers,'' IEEE Photon. J. \textbf{2}, 321-329 (2010).

\bibitem{Winful_88} S.S. Wang and H.G. Winful, ``Dynamics of phase-locked semiconductor laser arrays,'' Appl. Phys. Lett. \textbf{52}, 1774-1776 (1988).

\bibitem{Winful&Wang_88} H.G. Winful and S.S. Wang, ``Stability of phase locking in coupled semiconductor laser arrays,'' Appl. Phys. Lett. \textbf{53}, 1894-1896 (1988).

\bibitem{Winful_92} H.G. Winful, ``Instability threshold for an array of coupled semiconductor lasers,'' Phys. Rev. A \textbf{46}, 6093-6094 (1992).

\bibitem{LangKobayashi_80} R. Lang and K. Kobayashi, ``External Optical Feedback Effects on Semiconductor Injection Laser Properties,'' IEEE J. Quantum Electron. \textbf{QE-16}, 347-355 (1980).

\bibitem{Fischer_02} M. Peil, T. Heil, I. Fischer, and W. Elsaber, ``Synchronization of Chaotic Semiconductor Laser Systems: A Vectorial Coupling-Dependent Scenario,'' Phys. Rev. Lett. \textbf{88}, 174101 (2002).

\bibitem{Mandel_03} J. Javaloyes, P. Mandel, and D. Pieroux, ``Dynamical properties of lasers coupled face to face,'' Phys. Rev. E \textbf{67}, 036201 (2003).

\bibitem{Lenstra_05} H. Erzgraber, D. Lenstra, B. Krauskopf, E. Wille, M. Peil, I. Fischer, and W. Elsaber, ``Mutually delay-coupled semiconductor lasers: Mode bifurcation scenarios,'' Opt. Commun. \textbf{225}, 286-296 (2005).

\bibitem{Choquette_13} M.T. Johnson, D.F. Siriani, M.P. Tan, and K.D. Choquette, ``Beam steering via resonance detuning in coherently coupled vertical cavity laser arrays,'' Appl. Phys. Lett. \textbf{103}, 201115 (2013).

\bibitem{Choquette_15} S.T.M. Fryslie, M.T. Johnson, and K.D. Choquette, ``Coherence Tuning in Optically Coupled Phased Vertical Cavity Laser Arrays,'' IEEE J. Quantum Electron. \textbf{51}, 2600206 (2015).

\bibitem{Kominis_17b} Y. Kominis, V. Kovanis, and T. Bountis,  ``Controllable Asymmetric Phase-Locked States of the Fundamental Active Photonic Dimer,'' Phys. Rev. A \textbf{96}, 043836 (2017).

\bibitem{Adams_17} M.J. Adams, N. Li, B.R. Cemlyn, H. Susanto, and I.D. Henning, ``Effects of detuning, gain-guiding, and index antiguiding on the dynamics of two laterally coupled semiconductor lasers,'' Phys. Rev. A \textbf{95}, 053869 (2017).

\bibitem{Kovanis_97} A. Hohl, A. Gavrielides, T. Erneux, and V. Kovanis, ``Localized Synchronization in Two Coupled Nonidentical Semiconductor Lasers,'' Phys. Rev. Lett. \textbf{78}, 4745-4748 (1997).

\bibitem{Choquette_17mod} S.T.M. Fryslie, Z. Gao, H. Dave, B.J. Thompson, K. Lakomy, S. Lin, P.J. Decker, D.K. McElfresh, J.E. Schutt-Aine, and K.D. Choquette, ``Modulation of Coherently Coupled Phased Photonic Crystal Vertical Cavity Laser Arrays,'' IEEE J. Sel. Top. Quantum Electron. \textbf{23}, 1700409 (2017).

\bibitem{Kominis_19} Y. Kominis, K.D. Choquette, A. Bountis, and V. Kovanis, ``Antiresonances and Ultrafast Resonances in a Twin Photonic Oscillator,'' IEEE Photon. Journal \textbf{11}, 1500209 (2019).

\bibitem{Choquette_17} Z. Gao, S.T.M. Fryslie, B.J. Thompson, P. Scott Carney, and K.D. Choquette, ``Parity-time symmetry in coherently coupled vertical cavity laser arrays,'' Optica \textbf{4}, 323-329 (2017).

\bibitem{Choquette_18} Z. Gao, M.T. Johnson, and K.D. Choquette, ``Rate equation analysis and non-Hermiticity in coupled semiconductor laser arrays,'' J. Appl. Phys. \textbf{123}, 173102 (2018).

\bibitem{Kominis_17a} Y. Kominis, V. Kovanis, and T. Bountis, ``Spectral Signatures of Exceptional Points and Bifurcations in the Fundamental Active Photonic Dimer,'' Phys. Rev. A \textbf{96}, 053837 (2017).

\bibitem{Kominis_18} Y. Kominis, K.D. Choquette, A. Bountis, and V. Kovanis, ``Exceptional points in two dissimilar coupled diode lasers,'' Appl. Phys. Lett. \textbf{113}, 081103 (2018).

\bibitem{Erneux_book} T. Erneux and P. Glorieux, \textit{Laser Dynamics} (Cambridge University Press, 2010).
 
 \end{thebibliography}
\end{document}